\documentclass[onecolumn,preprint3]{aastex63}
\usepackage{lineno}
\usepackage{bbding}
\usepackage{amssymb,amsmath,bm}
\hypersetup{
  colorlinks      = {true},
  linkcolor       = {blue},
  citecolor       = {blue},
  urlcolor        = {blue},
}

\def\be{\begin{equation}}
\def\ee{\end{equation}}
\def\beq{\begin{eqnarray}}
\def\eeq{\end{eqnarray}}

\shorttitle{}
\shortauthors{Du et al.}
\begin{document}
\title{Lorentz Invariance Violation Limits from the Spectral Lag Transition of GRB~190114C}
\author[0000-0002-0986-218X]{Shen-Shi Du}
\affiliation{School of Physics and Technology, Wuhan University, Wuhan, Hubei 430072, China;}
\author{Lin Lan}
\affiliation{Department of Astronomy, Beijing Normal University, Beijing, China;}
\author{Jun-Jie Wei}
\affiliation{Purple Mountain Observatory, Chinese Academy of Sciences, Nanjing 210023, China;}\email{jjwei@pmo.ac.cn}
\affiliation{School of Astronomy and Space Sciences, University of Science and Technology of China, Hefei 230026, China;}
\author{Zi-Ming Zhou}
\affiliation{Guangxi Key Laboratory for Relativistic Astrophysics, School of Physical Science and Technology, Guangxi University, Nanning 530004, China;}
\author{He Gao}
\affiliation{Department of Astronomy, Beijing Normal University, Beijing, China;}\email{gaohe@bnu.edu.cn}
\author{Lu-Yao Jiang}
\affiliation{Guangxi Key Laboratory for Relativistic Astrophysics, School of Physical Science and Technology, Guangxi University, Nanning 530004, China;}
\author{Bin-Bin Zhang}
\affiliation{School of Astronomy and Space Science, Nanjing University, Nanjing 210093, China;}\email{bbzhang@nju.edu.cn}
\affiliation{Key Laboratory of Modern Astronomy and Astrophysics (Nanjing University), Ministry of Education, China}
\author{Zi-Ke Liu}
\affiliation{School of Astronomy and Space Science, Nanjing University, Nanjing 210093, China;}
\affiliation{Key Laboratory of Modern Astronomy and Astrophysics (Nanjing University), Ministry of Education, China}
\author{Xue-Feng Wu}
\affiliation{Purple Mountain Observatory, Chinese Academy of Sciences, Nanjing 210023, China;}
\affiliation{School of Astronomy and Space Sciences, University of Science and Technology of China, Hefei 230026, China;}
\author{En-Wei Liang}
\affiliation{Guangxi Key Laboratory for Relativistic Astrophysics, School of Physical Science and Technology, Guangxi University, Nanning 530004, China;}
\author{Zong-Hong Zhu}
\affiliation{School of Physics and Technology, Wuhan University, Wuhan, Hubei 430072, China;}
\begin{abstract}
The spectral lags of gamma-ray bursts (GRBs) have been viewed as the most promising probes of the possible violations of Lorentz invariance (LIV). However, these constraints usually depend on the assumption of the unknown intrinsic time lag in different energy bands and the use of a single highest-energy photon. A new approach to test the LIV effects has been proposed by directly fitting the spectral lag behavior of a GRB with a well-defined transition from positive lags to negative lags. This method simultaneously provides a reasonable formulation of the intrinsic time lag and robust lower limits on the quantum-gravity energy scales ($E_{\rm QG}$).
In this work, we perform a global fitting to the spectral lag data of GRB~190114C by considering the possible LIV effects based on a Bayesian approach. We then derive limits on $E_{\rm QG}$ and the coefficients of the Standard Model Extension. The Bayes factors output in our analysis shows a very strong evidence for the spectral-lag transition in GRB~190114C.
Our constraints on a variety of isotropic and anisotropic coefficients for LIV are somewhat weaker than existing bounds,
but they can be viewed as comparatively robust and have the promise to complement existing LIV constraints.
The observations of GRBs with higher-energy emissions and higher temporal resolutions will contribute to a better formulation of the intrinsic time lag and more rigorous LIV constraints in the dispersive photon sector.

\end{abstract}
\keywords{astroparticle physics --- gamma-ray burst: individual (GRB~190114C) --- gravitation}

\section{Introduction}\label{Sec:introduction}
Spectral lag, referring to the different arrival times for the photons with different energies, is a common feature in gamma-ray burst (GRB) pulses observed during the prompt emission \citep{Norris86ApJ,Norris00ApJ, Cheng95AAP,Band97ApJ,Chen05ApJ,Peng07CJAA}. Observationally, most of the GRB pulses are dominated by positive lags (i.e., the soft photons lag behind the hard photons), and a small fraction of pulses show zero and even negative lags (e.g, \citealt{Norris00ApJ, Yi06MNRAS, Wei2017ApJL}). It was suggested that the high-latitude emission ``curvature effect" of a relativistic jet can plausibly account for the intrinsic origin of (positive) spectral lags (e.g., \citealt{Ioka01ApJL,Shen05MNRAS,Lu06MNRAS,Shenoy13ApJ}). \cite{Zhang09ApJ} pointed out that the main difficulty of the curvature effect scenario is that the flux levels at different energy bands are significantly lower than those observed. \cite{Uhm&Zhang16} further showed that there would be essentially unnoticeable time lags given rise from the high-latitude emission of a spherical fireball and the reproduced properties of light curves are not compatible with the observations. Instead, they established a physical model invoking a spherical shell rapidly expands in the bulk-accelerating region, which is suggested to account for the spectral lags more reasonably \citep{Wei2017ApJL,Shao17ApJ,Lu18ApJ,Uhm2018ApJ}.
On the other hand, \cite{Lu18ApJ} found that the spectral lag is closely related to the spectral evolution within 92 pulses from 84 GRBs observed by the Fermi Gamma-ray Burst Monitor (GBM). These researches motivate the work by \cite{Du2019ApJ}, who investigated the detailed relationship of the spectral lag with spectral evolution by using a phenomenological model with different spectral evolution patterns. Their results can naturally reproduce the negative lags observed in GRB~160625B for more generic physical models.

The candidate quantum-gravitational (QG) effect in the particle-physics field provides an explanation of the negative lags.
One such effect is the so-called Lorentz invariance violation (LIV).
There is great interest in the expectation that the Lorentz invariance breaks presumably at the energy on the \emph{Planck} scale (i.e., $E_{\rm QG}\approx E_{\rm pl}=\sqrt{hc^5/2\pi G}\simeq 1.22\times 10^{19}$~GeV; \citealt{Amelino-Camelia1998Natur, Mattingly2005LRR, Amelino-Camelia2013LRR}).
Such type of QG effects may appear as the high-energy photons interact with the `foam' on short time and small scales \citep{Amelino-Camelia1997IJMPA}, resulting in the in vacuo photon dispersion (i.e., an energy-dependent speed of light).
In such scenarios, the high-energy photons propagate more slowly than the low-energy photons in vacuum \citep{Amelino-Camelia1998Natur}.
Thus, the discrepant arrival time of photons with different energies emitted from the same astrophysical source allows us to constrain the quantum-gravity energy scale $E_{\rm QG}$ \citep{Amelino-Camelia1998Natur,Mattingly2005LRR,Ellis2013APh}.

GRBs are characterized by long distances, short spectral lags, and very energetic emissions, which make them the current most powerful probes to the energy scales of LIV in the dispersive photon sector \citep{Mattingly2005LRR, Abdo2009Sci, Abdo2009Natur, Wei2017ApJL,Ellis2019PhRvD}. The bounds on LIV effects have been obtained by analyzing the rough time lag of a single high-energy photon detected in some bursts or the arrival times of sharp features of intensity profile with statistical samples of GRBs  (\citealt{Amelino-Camelia1998Natur, Schaefer1999PhRvL, Ellis2003A&A, Ellis2006APh, Boggs2004ApJ,Abdo2009Sci, Abdo2009Natur, Xiao2009PhRvD, Chang2012APh, Nemiroff2012PhRvL, Zhang2015APh}).
The most stringent lower limit on the linear LIV leading term of vacuum dispersion with Taylor expansion comes from the studies of the effects on the time of flight by analyzing the arrival time delay between a 31~GeV photon and low-energy photons from short GRB~090510, with $E_{\rm QG,1}>(1-10) \times E_{\rm pl}$ \citep{Abdo2009Sci, Vasileiou2013PhRvD}.
For the quadratic LIV, the strictest lower limit has resulted from the observations of active galactic nuclei Markarian 501 (Mrk 501) by the H.E.S.S. telescope, i.e., $E_{\rm QG,2}>8.5 \times 10^{10}$~GeV ($E_{\rm QG,2}>7.8\times 10^{11}$~GeV) from the arrival time delays of multi-TeV photons (from the modifications to the interactions between very high energy gamma-rays and extragalactic background light photons; \citealt{Abdalla2019ApJ}).
Recently, the TeV gamma-rays detected from GRB~190114C by the Major Atmospheric Gamma Imaging Cherenkov (MAGIC) telescopes yield a competitive lower limit $E_{\rm QG,2}>6.3\times 10^{10}$~GeV ($E_{\rm QG,2}>5.6\times 10^{10}$~GeV) in the subluminal (superluminal) scenario \citep{Acciari2020PhRvL.125b1301A}.
Actually, these lower constraints correspond to no significant LIV effects but suffer from some conservative assumptions.
For instance, in \cite{Acciari2020PhRvL.125b1301A}, it is not a multi-wavelength fit in investigating effects on the time of flight; there is no comparison in investigating the likelihood of synchrotron-self-Compton versus a base synchrotron explanation; the parameters of intrinsic spectral and temporal emissions are assumed. 
Furthermore, distinguishing an intrinsic time delay at the source from a delay caused by the propagation effects in a vacuum is a critical obstacle. \cite{Ellis2006APh} first considered the possible intrinsic time lag ($\Delta t_{\rm int}$) by fitting the spectral lags extracted from broad light curves observed in different energy bands with a linear regression model in which the slope is related to the LIV effects and the intercept corresponds to the $\Delta t_{\rm int}$ in the observer. This approach directs to weak evidence for LIV by assuming that all GRBs have the same intrinsic time lag. Thus far, it is necessary to constrain LIV by using the detailed spectral lag measurements from high-quality and high-energy light curves from GRBs.

Recently, \cite{Wei2017ApJL} used the spectral lag data between the light curves observed in different energy bands from GRB~160625B to put robust constraints on the LIV effects (see also \citealt{Pan2020ApJ}).
This method, for the first time, gives a reasonable formulation for the intrinsic (positive) time lags, and the $E_{\rm QG}$ can be constrained if a transition from positive lags to negative lags appears in high-energy range\footnote{The potential LIV effect that is expected to appear on QG scales could lead to the energy-dependent time delay in the arrival times of observed photons, it would influence the observed temporal and spectral features (e.g., the spectral evolution) of GRBs and contribute to the entire observed spectral lags. As shown by \cite{Du2019ApJ}, different spectral evolution patterns can produce different evolving behaviors of spectral lags in GRB pulses, which would inherently include the contributions of the LIV effects if which exist.}. In this way, the observed arrival time delay between light curves in different energy bands are comprised of two contributions\footnote{In fact, for cosmic transient sources (e.g., GRB) there are three other effects attributed to the time delay between photons with different energies \citep{Gao2015ApJ}: (1) the potential time delay from the special-relativistic effects in the case that photons have non-zero rest mass; (2) the time delay from the dispersion by the free electrons in the line of sight and (3) the gravitational potential integrated from the emission site to the Earth if Einstein's equivalence principle is violated. The contributions of those three effects can be negligible for the photons emitted from GRBs, as shown by \citep{Gao2015ApJ} and \citep{Wei2015PhRvL}.}
\begin{equation}\label{Eq:t_obs}
\Delta t_{\rm obs}=\Delta t_{\rm LIV}+\Delta t_{\rm int,z},
\end{equation}
where $\Delta t_{\rm LIV}$ is the observed time lags induced by the LIV effects and $\Delta t_{\rm int,z}=\Delta t_{\rm int}(1+z)$ is the intrinsic time lags corrected by the cosmological time dilation.
In this work, we show that, in GRB~190114C, the spectral lag monotonically increases with the photon energy and approximately saturates at a certain photon energy. In particular, we show decisive evidence that there is a transition from positive lags to negative lags well measured at $\sim0.7$~MeV, lower than that in GRB~160625B with one order of magnitude ($\sim8$~MeV, \citealt{Wei2017ApJL}). Thus, GRB~190114C is another case that allows us to constrain the quantum-gravity energy scale by directly fitting its observed spectral lag data.

The contents of this paper are arranged as follows. In Section~\ref{Sec:Sec2}, we describe the observed properties of GRB~190114C. We present our method and constraint results in Section~\ref{Sec:Sec3}. Our conclusion and discussions are presented in Section~\ref{Sec:Sec4}.
The flat $\Lambda$CDM cosmology (radiation density is ignored) with parameters $\Omega_{\rm m,0}=0.315$ ($\Omega_{\Lambda,0}=1-\Omega_{\rm m,0}$) and $H_0=67.36$~km$\cdot$s$^{-1}\cdot$Mpc$^{-1}$ \citep{Planck2018arXiv180706209P} is adopted throughout this work.


\section{The Observed Properties of GRB~190114C}\label{Sec:Sec2}
On 14 January 2019 at 20:57:03 universal time, GRB~190114C was first triggered by the GBM onboard the Fermi Gamma-Ray Space Telescope \citep{Hamburg2019GCN.23707....1H}, and the Burst Alert Telescope (BAT) onboard the Neil Gehrels Swift Observatory located this burst with coordinates (J2000) right ascension, ${\rm R. A.}=54.5^{\circ}$, and declination, ${\rm Dec} =-26.9^{\circ}$ \citep{Gropp2019GCN.23688....1G}. 
BAT and GBM measured the duration time of this burst in terms of $T_{90}$  (the time interval containing 90\% of the total photon counts)  to be $\sim362$~s (in 15$-$350~keV band) and $\sim116$~s (in 50$-$300~keV band), respectively. Spectroscopic observations reveal absorption features consistent with Ca H\&K, Na ID, Mg II and Mg I at a redshift of $z=0.4245\pm0.0005$ \citep{Selsing2019GCN.23695....1S, Kann2019GCN.23710....1K}. Remarkably, MAGIC detected the very-high-energy (VHE) emissions up to at least 1~TeV from GRB 190114C about 50~s after the \emph{Swift} trigger \citep{MAGIC2019Natur}. \emph{Fermi} Large Area Telescope (LAT) also detected this source, lasting until $\sim$150~s after GBM trigger \citep{Kocevski2020AAS}.

\begin{figure}
\begin{center}
\begin{tabular}{ccc}
\includegraphics[angle=0,scale=0.35]{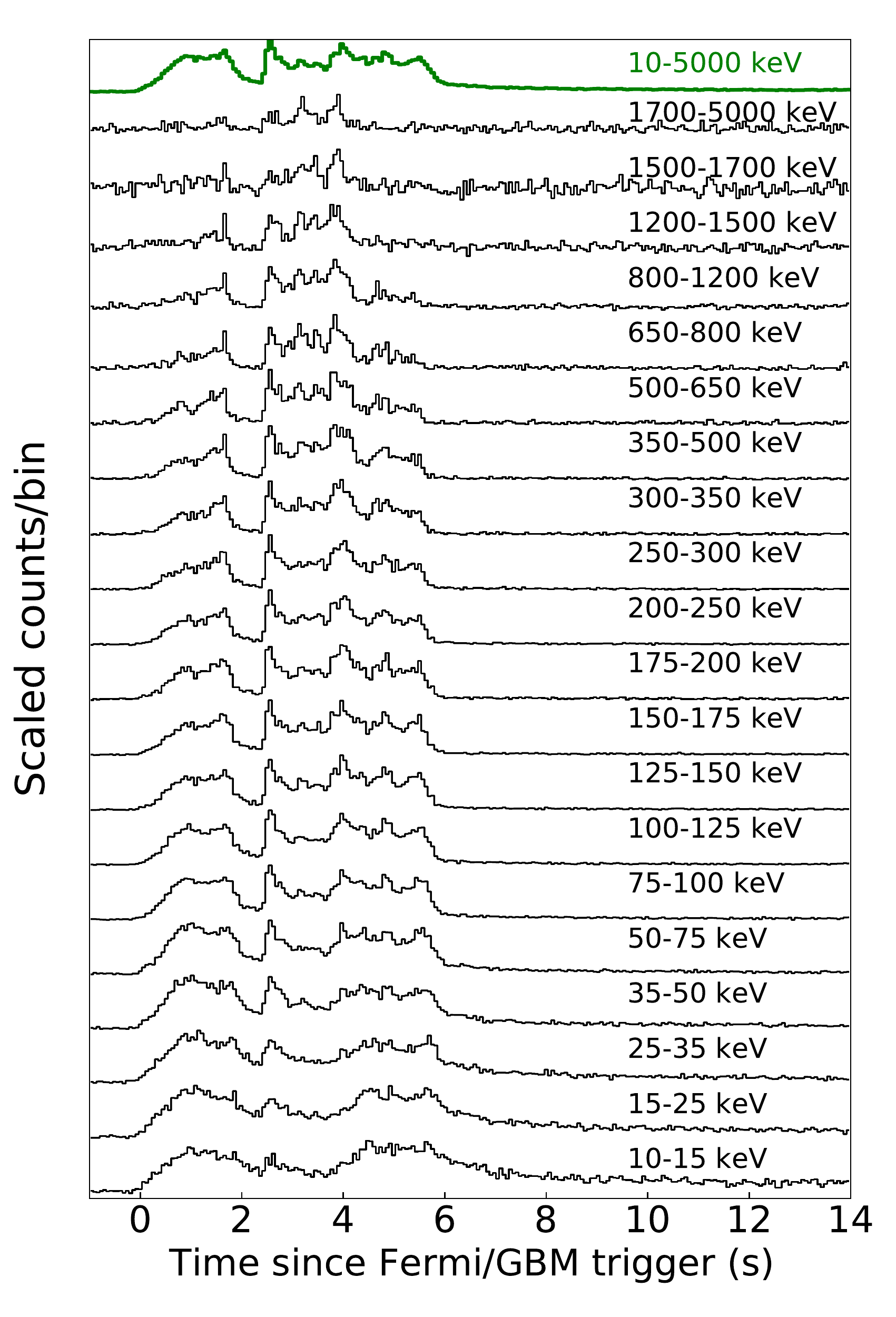}
\end{tabular}
\end{center}
\caption{Energy-dependent light curves of GRB~190114C observed by Fermi/GBM. The top thick green line shows the full-range ($10-5000$~keV) light curve.
}\label{MyFigA}
\end{figure}

Figure~\ref{MyFigA} shows the \emph{Fermi}-GBM light curves of GRB~190114C during $-1-14$~s (the main pulses) observed in different energy bands.
With the high photon statistics for the bright light curves of GRB~190114C in amply populated energy bands shown in Figure~\ref{MyFigA}, the time lags between the light curves observed in different energy bands are calculated by employing the cross-correlation function (CCF) method. We direct the interested readers to \cite{Ukwatta2010ApJ} and \cite{ZhangBB2012ApJ} for more details of the CCF method. We adopt the arrival time of the time series in the lowest energy band ($10-15$~keV) as a reference, and the time lags between this energy band and any other high-energy bands are shown in Figure~\ref{MyFigB} and tabulated in Table~\ref{MyTabA}. One can see that the spectral lag monotonically increases with energy and saturates at about 60~keV. In particular, there is an obvious transition from the positive lags to negative lags at around 0.7~MeV, which is analogous to (but the transition point is about one order of magnitude smaller than) that observed in GRB~160626B ($\sim$8~MeV) \citep{Wei2017ApJL}. Considering the negative lags that the LIV involves, this behavior makes GRB~190114C the another case that allows us to constrain the energy scales of LIV effects by directly fitting its spectral lag data.


\section{LIV constraints}\label{Sec:Sec3}
\subsection{Formulating the Intrinsic Energy-dependent Time Lag}\label{Sec:Intlag}
We formulate that, in the observer frame, the relation of the intrinsic time lag and the photon energy $E$ (adopted as the median value of each energy band) in the form of a power-law function:
\begin{equation}\label{Eq:tint_Formula}
\Delta t_{\rm int,z} (E)=\tau \left[\left( \frac{{\mathcal E}_0}{\rm 1~keV} \right)^{-\alpha} - \left(\frac{E}{\rm 1~keV}\right)^{-\alpha}\right],
\end{equation}
where ${\mathcal E}_0=12.5$~keV is the median value of the lowest reference energy band (10--15~keV). The coefficient $\tau$ and index $\alpha$ are the free parameters to be simultaneously constrained together with the LIV energy scale $E_{\rm QG}$. Specifically, $\tau>0$ and $\alpha>0$ corresponds to that the intrinsic time lag has a positive dependence on $E$ (i.e., higher energy emission arrives earlier in the observer).

\subsection{Vacuum Dispersion with Taylor Expansion}\label{Sec:LIVlag}
Assuming that General Relativity is correct in the assumptions of the constant speed of light and the validity of Lorentz invariance, the energy-dependent velocity of light would be $E^2=p^2c^2$, where $p$ is the momentum of the photon. In QG scenarios, the introduction of LIV leads to the modifications to the photon dispersion relation (for the photons with energy $E\ll E_{\rm pl}$; \citealt{Amelino-Camelia1998Natur})
\begin{equation}
E^2\simeq p^2c^2\left[1-\sum_{n=1}^{\infty}s_{\pm}\left(\frac{E}{E_{\rm QG}}\right)^n\right],
\end{equation}
and the photon propagation speed can be expressed as
\begin{equation}\label{Eq:v_ph}
v (E)=\frac{\partial E}{\partial p}\simeq c\left[1-\sum_{n=1}^{\infty}s_{\pm}\frac{n+1}{2}\left(\frac{E}{E_{\rm QG,n}}\right)^n\right],
\end{equation}
where $E_{\rm QG}$ denotes the QG energy scale,
the linear ($n=1$) or quadratic ($n=2$) term is generally adopted as the $n$th order Taylor expansion of the leading LIV term, and $s_{\pm}=\pm 1$ represents the sign of the LIV effect corresponding to the subluminal ($s_{\pm}=+1$) or superluminal  ($s_{\pm}=-1$) scenario (i.e., $s_{\pm}=+1$ or $s_{\pm}=-1$ stands for a decrease or an increase in photon group velocity with an increasing photon energy).
Thus, $s_{\pm}=+1$ would be the case that higher-energy photons propagate more slowly relative to the lower-energy photons in a vacuum. This gives the LIV-induced negative time lags. Thus we only consider the case of $s_{\pm}=+1$ in this work.

Eq.~(\ref{Eq:v_ph}) shows that two photons with different energies (denoted by $E$ and ${\mathcal E}_0$, $E>{\mathcal E}_0$) emitted from the same source would propagate with different velocities, thus arrive on Earth at different times. Considering the cosmological time dilation, the LIV-induced time delay between high-energy and low-energy photons is given such that \citep{Zhang2015APh}
\begin{equation}\label{Eq:t_LIV}
\Delta t_{\rm LIV}(E)=t_{\rm l}-t_{\rm h}\approx -\frac{1+n}{2}\frac{E^n-{\mathcal E}_0^{n}}{E^n_{{\rm QG},n}}\int_{0}^{z}\frac{({1+\xi})^nd\xi}{H(\xi)},
\end{equation}
where $H(z) = H_0[\Omega_{\rm m,0}(1+z)^3+\Omega_{\Lambda,0}]^{1/2}$ is the expansion history of the Universe, $t_{\rm h}$ and $t_{\rm l}$ are the observed arrival times of the photons with energies $E$ and ${\mathcal E}_0$, respectively.

\subsection{Vacuum Dispersion with Standard Model Extension}\label{Sec:LIVlag_SME}
Alternatively, at attainable energies, the LIV effects can be described in the framework of standard-model extension (SME; \citealt{1997PhRvD..55.6760C,1998PhRvD..58k6002C,2004PhRvD..69j5009K}) based on effective field theory \citep{Kosteleck1995PhRvD..51.3923K}. In this scenario, the corresponding modified dispersion relations are solved as \citep{Kosteleck2008ApJ...689L...1K, Kosteleck2009PhRvD..80a5020K, Wei2017ApJ...842..115W}
\begin{equation}\label{Eq:SME_E}
E(p) \simeq (1-\varsigma^0 \pm \sqrt{(\varsigma^0)^2+(\varsigma^2)^2+(\varsigma^3)^2} )\;p,
\end{equation}
with $p$ being the photon momentum as a function of the frequency. The symbols $\varsigma^0$, $\varsigma^1$, $\varsigma^2$, and $\varsigma^3$ represent the momentum- and direction-dependent combinations of coefficients for Lorentz violation. These quantities can be decomposed through spin-weighted spherical harmonics, ${}_{s}{Y}_{jm}(\hat{\textbf{\emph{n}}})$, of spin weight, $s$, with a well-understood set \citep{Newman1966JMP.....7..863N,Goldberg1967JMP.....8.2155G}:
\begin{equation}\label{Eq:SME_sigma}
\begin{aligned}
  \varsigma^0 &=
\sum_{djm}p^{d-4} {}_{0}Y_{jm}(\hat{\textbf{\emph{n}}})c_{(I)jm}^{(d)},
\\
  \varsigma^1 \pm i\varsigma^2 &=
\sum_{djm}p^{d-4} {}_{\mp2}{Y}_{jm}(\hat{\textbf{\emph{n}}})
\left(k_{(E)jm}^{(d)} \mp ik_{(B)jm}^{(d)}\right),
\\
  \varsigma^3 &=
\sum_{djm}p^{d-4} {}_{0}Y_{jm}(\hat{\textbf{\emph{n}}})k_{(V)jm}^{(d)},
\end{aligned}
\end{equation}
where $\hat{\textbf{\emph{n}}}$ is a unit vector pointing towards the source, $d$ is the arbitrary mass dimension (in natural units of $\hbar = c = 1$) for each term in the SME  Lagrange density with Lorentz-violating operators, and $jm$ provide the eigenvalues of total angular momentum for each combination of coefficient. Generally, the standard spherical polar coordinates ($\theta, \phi$) are adopted for $\hat{\textbf{\emph{n}}}$ and defined in a Sun-centered celestial-equatorial frame \citep{Kosteleck2002PhRvD..66e6005K}, with $\theta =90^{\circ}- {\rm Dec}$ and $\phi={\rm R. A.}$.

The decomposition defined above characterizes all types of Lorentz violations for vacuum propagation with four sets of spherical coefficients: $c_{(I)jm}^{(d)}$, $k_{(E)jm}^{(d)}$, and $k_{(B)jm}^{(d)}$ are the coefficients for even $d$ and control CPT-even effects; $k_{(V)jm}^{(d)}$ are the coefficients for odd $d$ and govern CPT-odd effects.
For instance, in the isotropic limit ($j=m=0$), the group-velocity defect for photons is \citep{Wei2017ApJ...842..115W}
\begin{equation}
\delta v_g(E) \simeq
\frac{1}{\sqrt{4\pi}} \sum_d (d-3) E^{d-4}
\big(- c_{(I)00}^{(d)} \pm k_{(V)00}^{(d)} \big)\;.
\end{equation}
Birefringence results when the usual degeneracy among polarizations is broken, for which at least one of $k_{(E)jm}^{(d)}$, $k_{(B)jm}^{(d)}$, and $k_{(V)jm}^{(d)}$ is nonzero.
Here, we focus on the direction-dependent dispersion constraints on non-birefringent Lorentz-violating effects, thereby the only
coefficients in Eq.~(\ref{Eq:SME_sigma}) are $c_{(I)jm}^{(d)}$.
Tests for vacuum dispersion seek energy-dependent velocity of light, corresponding to the cases of even $d>4$ and nonzero values for $c_{(I)jm}^{(d)}$.
Setting to zero the coefficients for birefringent propagation, the group-velocity defect can be given by
\begin{equation}
\delta v_g (E) = - \sum_{djm} (d-3)
E^{d-4} \, {}_{0}Y_{jm}(\theta, \phi) c_{(I)jm}^{(d)}\;,
\end{equation}
in which all the spherical coefficients can be taken to be constants at ($\theta, \phi$) in the Sun-centered frame.
This leads to an arrival-time difference for two photons with different energies $E_{\rm h}>{\mathcal E}_0$ simultaneously emitted from the same source at redshift $z$, that is \citep{Kosteleck2008ApJ...689L...1K,Wei2017ApJ...842..115W}
\begin{equation}\label{Eq:SME_tLIV}
\begin{aligned}
\Delta t_{\rm LIV} (E) &=t_{\rm l}-t_{\rm h}
\\
&
\approx
-(d-3)\left(E_{\rm h}^{d-4} - {\mathcal E}_0^{d-4}\right)
\int_0^{z}\frac{(1 + \xi)^{d-4}}{H(\xi)} {\rm d} \xi
\sum_{jm} {}_{0}Y_{jm}(\theta, \phi) c_{(I)jm}^{(d)}
\;,
\end{aligned}
\end{equation}
where the coefficients $ c_{(I)jm}^{(d)}$ can be either positive or negative, leading to a decreasing or an increasing velocity of light with photon energy.
Thus, a positive $\sum_{jm} {}_{0}Y_{jm}(\theta, \phi) c_{(I)jm}^{(d)}$ (denoted by $\rm LIV_{SME}$) would imply a negative spectral lag contributed by Lorentz violation, which is the case we consider in the following discussions.

\subsection{Results: Bayesian analysis}\label{Sec:constraints}
\begin{figure}
\centering
\includegraphics[angle=0,scale=0.5]{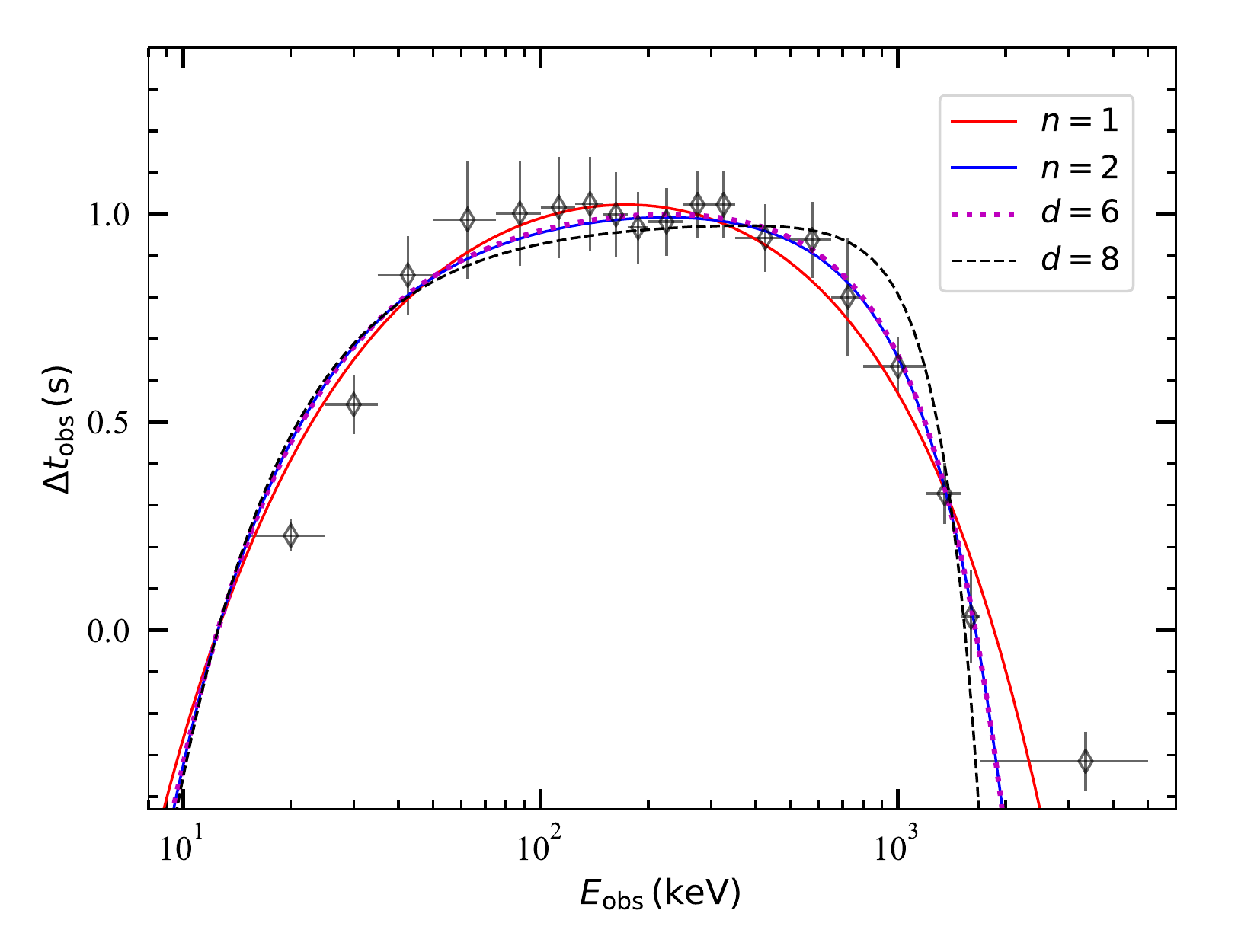}
\caption{The observed time lags between the softest energy band (10--15~keV) and any other high energy bands.
The color lines correspond to the median values of estimated constraints for LIV effects parametrized by vacuum dispersion in Taylor expansion and SME (see Table~\ref{MyTabB}).
}\label{MyFigB}
\end{figure}

With the observed time lags between the reference energy band ($10-15$~keV) and 19 higher energy bands (see Table~\ref{MyTabA}), we simultaneously fit the free parameters from Eqs.~(\ref{Eq:t_obs}), (\ref{Eq:tint_Formula}), and (\ref{Eq:t_LIV}) (or Eq.~(\ref{Eq:SME_tLIV})) with a Beyesian approach.
Given the observed spectral lag data ${\bm D}_{\rm \tau}$ from GRB~190114C (see Figure~\ref{MyFigB}) and some prior knowledge about the hypothetical models (parameters are denoted by the vector ${\bm \theta}$, see below), the posterior probability distribution of the free parameters, i.e., the Bayesian \emph{posterior} is expressed with \citep{Trotta2008ConPh}
\begin{equation}\label{Eq:Bayes}
p({\bm \theta | {\bm D}_{\rm \tau}})=\frac{{\mathcal L}({\bm \theta}; {\bm D}_{\rm \tau}) p (M({\bm \theta}))}{\int {{\mathcal L}({\bm \theta}; {\bm D}_{\rm \tau}) p(M({\bm \theta}))} d{\bm \theta}}.
\end{equation}
${\mathcal L}({\bm \theta}; {\bm D}_{\rm \tau})=p({\bm D}_{\rm \tau}|M({\bm \theta} ))$ is the \emph{likelihood} of the spectral lag data conditional on the knowledge of LIV and intrinsic time lag models, which can be written as
\begin{equation}\label{Eq:likelihood}
\mathcal{L} ({\bm \theta}; {\bm D}_{\rm \tau})=\prod_{i=1}^{19}\frac{1}{\sqrt{2\pi [\sigma_{\Delta t_{\rm obs,i}}^2+ \sigma^2_{\Delta t_{\rm mod,i}} ({\bm \theta})}]}{\rm exp}\left\{-\frac{[{\Delta t_{\rm obs,i}}- \Delta t_{\rm mod,i}({\bm \theta})]^2}{2[\sigma^2_{\Delta t_{\rm obs,i}}+ \sigma^2_{\Delta t_{\rm mod,i}} ({\bm \theta})^2]}\right\},
\end{equation}
where $\Delta t_{\rm mod}$ is the theoretical spectral lags including the intrinsic and LIV-induced time lags and its error, $\sigma_{\Delta t_{\rm mod}}$, is propagated from Eqs.~(\ref{Eq:tint_Formula}) and (\ref{Eq:t_LIV}) (or Eq.~(\ref{Eq:SME_tLIV})) via $\sigma_{\Delta t_{\rm mod}} = [\Delta {\dot t}_{\rm int,z}(E)+ {\Delta {\dot t}_{\rm LIV}}(E)]\;\sigma_{\rm E}$, with $\sigma_{\rm E}=(E_{\rm max}-E_{\rm min})/2$ being the statistical error of photon energy in the energy band ($E_{\rm min}, E_{\rm max}$).
The $p(M({\bm \theta}))$ is the Bayesian \emph{prior}, and a uniform \emph{prior} is adopted with large (and acceptable) ranges in this work.
The uniform priors for spherical coefficients in Eq.~(\ref{Eq:SME_tLIV}) are somewhat confined with LIV$_{\rm SME}>0$, corresponding to the cases of negative time lags due to LIV effects.
The denominator in Eq.~(\ref{Eq:Bayes}) represents the Bayesian \emph{evidence}, which is an integration over the whole parameter space of $\bm \theta$.
By implementing the dynamic nested sampling method {\tt Dynesty} \citep{Speagle2020MNRAS.493.3132S},\footnote{\url{https://github.com/joshspeagle/dynesty}} we can effectively and simultaneously estimate the \emph{posterior} and \emph{evidence} while optimizing the likelihood function given by Eq.~(\ref{Eq:likelihood}) to obtain the \emph{posterior} probability distributions of such free parameters. 

\begin{figure}
\centering
\includegraphics[angle=0,scale=0.35]{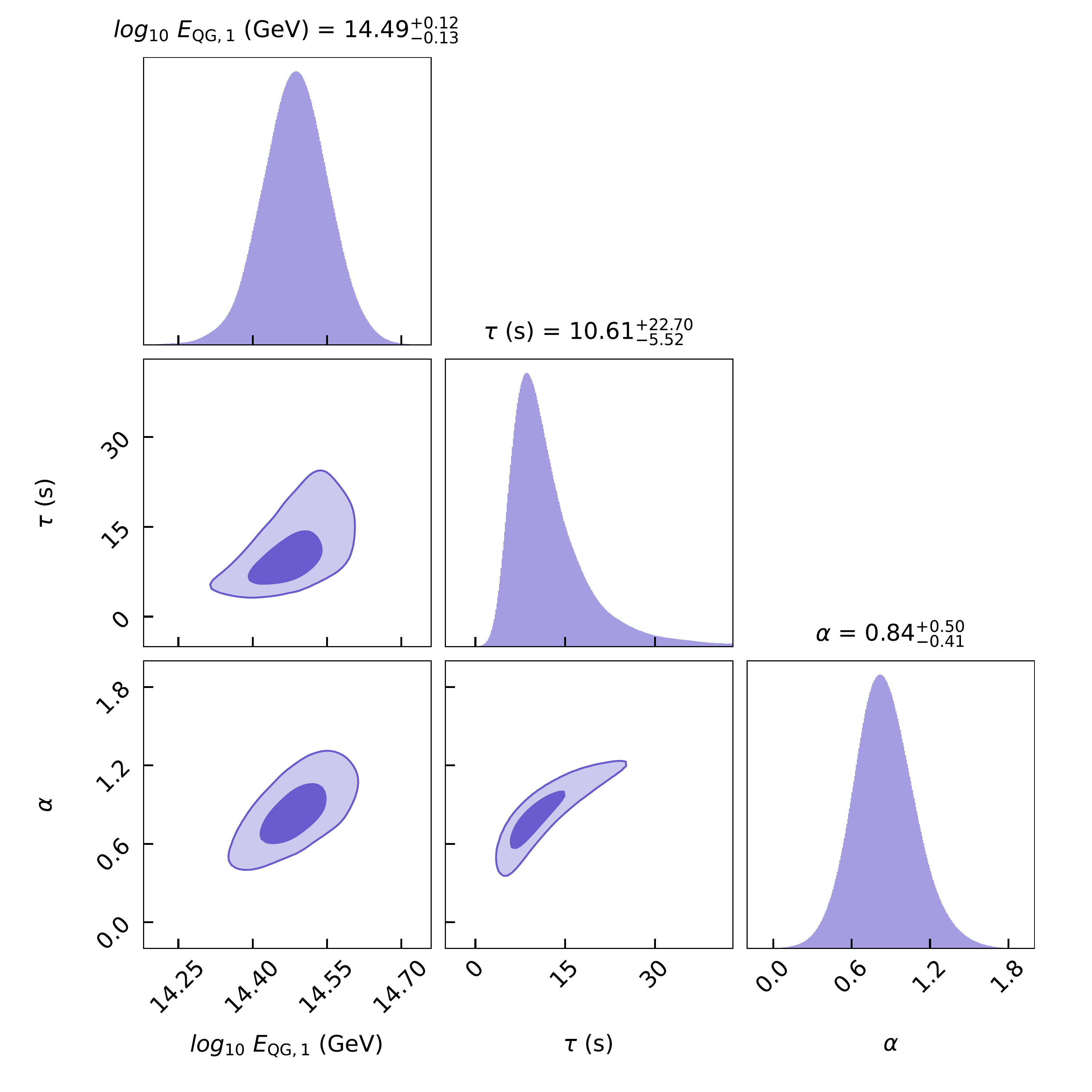}
\includegraphics[angle=0,scale=0.35]{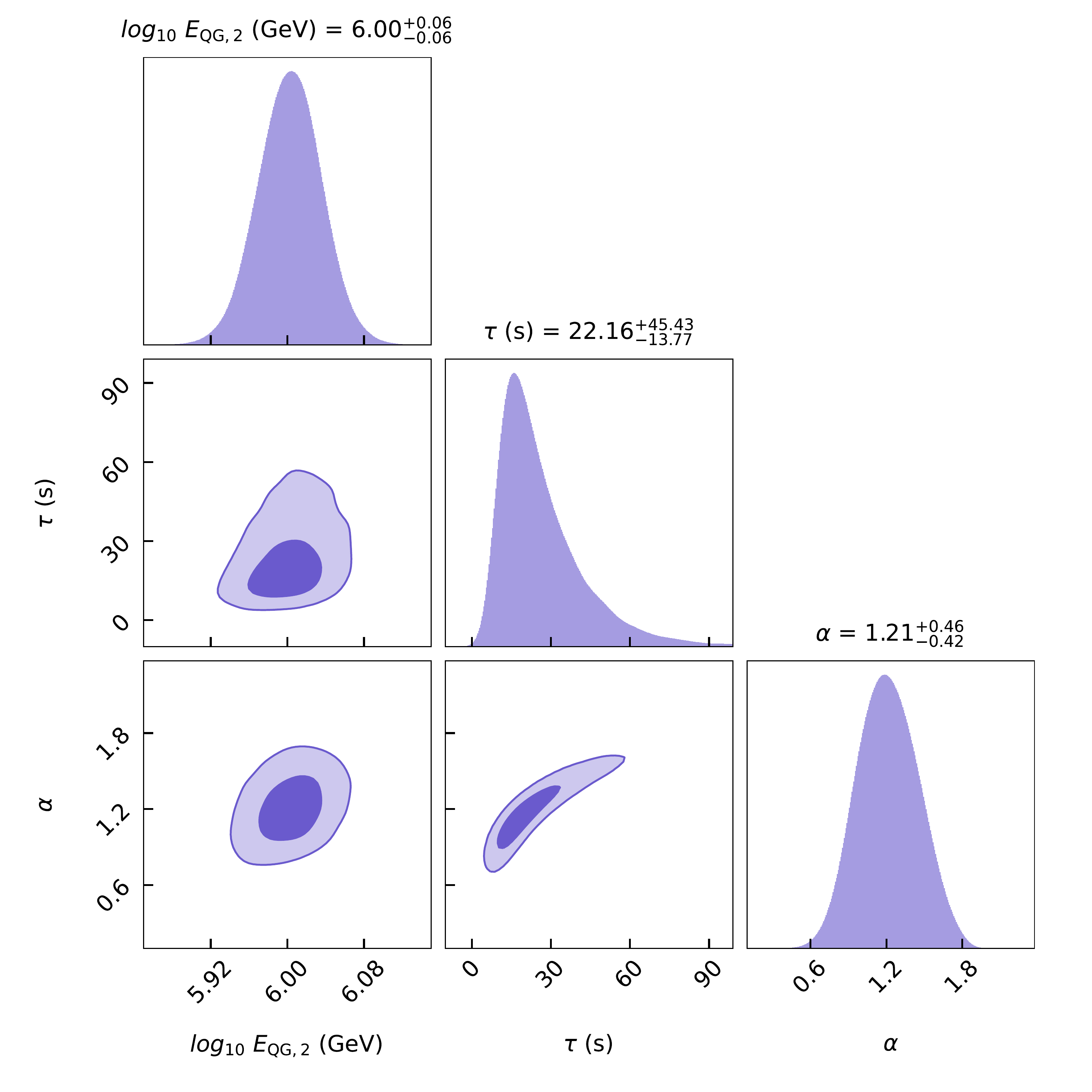}
\caption{Marginalized one- and two-dimensional posterior probability distribution contours at 1$\sigma$ and 2$\sigma$ confidence levels obtained in the case of the vacuum dispersion parametrized by Taylor series expansion. Parameter constraints shown on the top and bottom panels correspond to the cases of  linear ($n=1$) and quadratic ($n=2$) LIV models, respectively.  The titles shown on the top of each subplot represent the 2$\sigma$ credible regions. 
}\label{MyFigC}
\end{figure}

With the LIV-induced time delay for Taylor expansion, the free parameters to be constrained are ${\bm \theta}=\{E_{\rm QG}, \tau, \alpha\}$. Hereafter the constraints are taken within 2$\sigma$ confidence levels unless noted otherwise. The marginalized one- and two-dimensional \emph{posterior} probability distributions output from dynamic nested sampling code are drawn in the corner plots in Figure~\ref{MyFigC}. From the top panel of Figure~\ref{MyFigC}, one can see that the limits for the linear ($n=1$) LIV model are $\log_{10} (E_{\rm QG,1}/{\rm GeV})=14.49^{+0.12}_{-0.13}$, $\tau(/{\rm s})=10.61^{+20.70}_{-5.52}$, and $\alpha=0.84^{+0.50}_{-0.41}$ (see also Table~\ref{MyTabB}). The median values give a reduced chi-squared $\chi^2_{\rm dof}=0.52$ with sixteen degrees of freedom.
For the quadratic ($n=2$) LIV case (bottom panel), we obtain $\log_{10} (E_{\rm QG,2}/{\rm GeV})=6.0^{+0.06}_{-0.06}$, $\tau(/{\rm s})=22.16^{+45.43}_{-13.77}$, and $\alpha=1.21^{+0.46}_{-0.42}$. Using the median values for the quadratic LIV, we obtain $\chi^2_{\rm dof}=0.46$ with sixteen degrees of freedom.
Figure~\ref{MyFigB} displays the theoretical curves of spectral lags corresponding to the above median values along with the confidence bands of 2$\sigma$ error bars. The results from the 2$\sigma$ confidence levels yield lower limits on the LIV energy scale $E_{\rm QG,1}\ge 2.23\times 10^{14}$~GeV for the linear ($n=1$) LIV term and $E_{\rm QG,2}\ge 0.87\times 10^6$~GeV for the quadratic ($n=2$) LIV case, respectively.

\begin{figure}
\centering
\includegraphics[angle=0,scale=0.35]{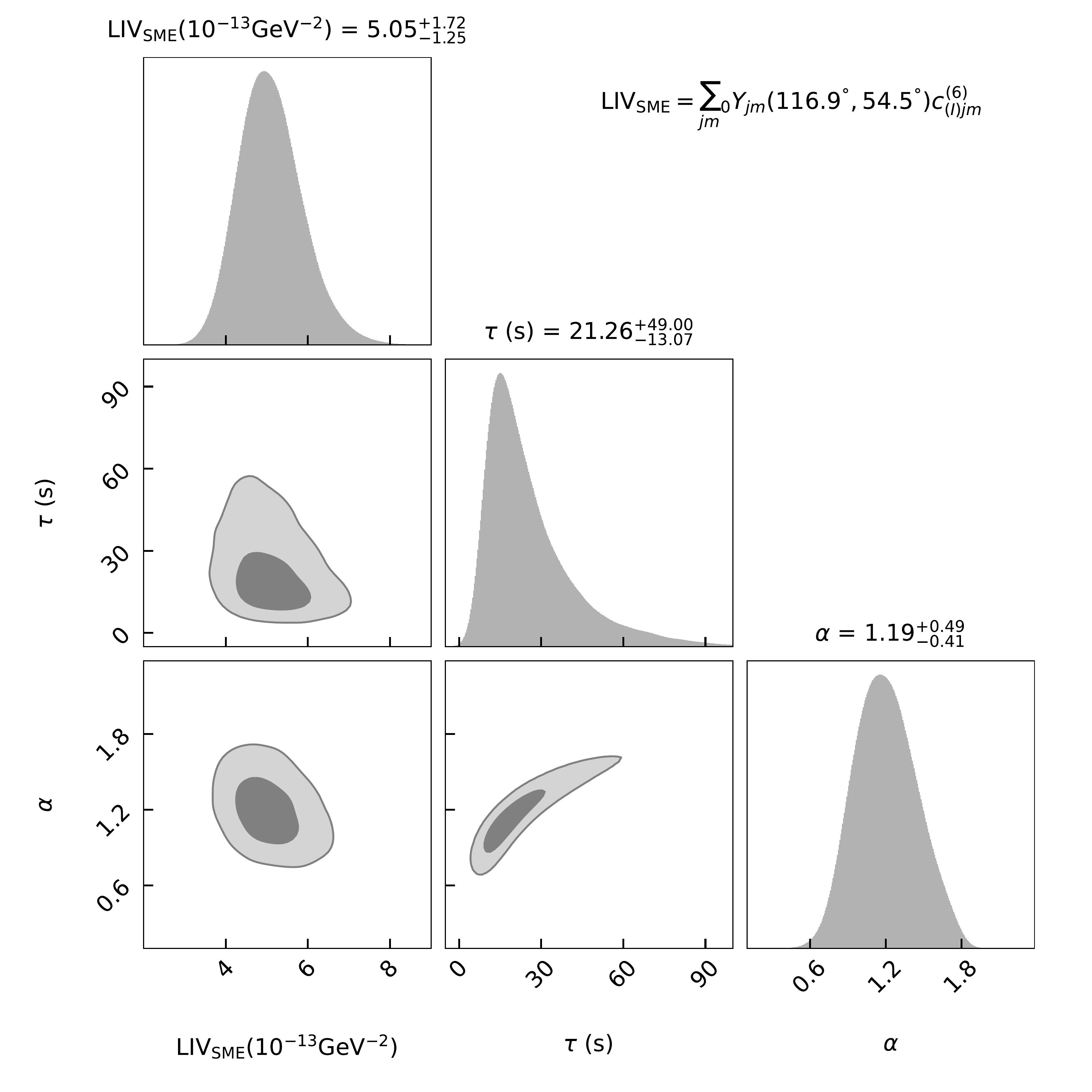}
\includegraphics[angle=0,scale=0.35]{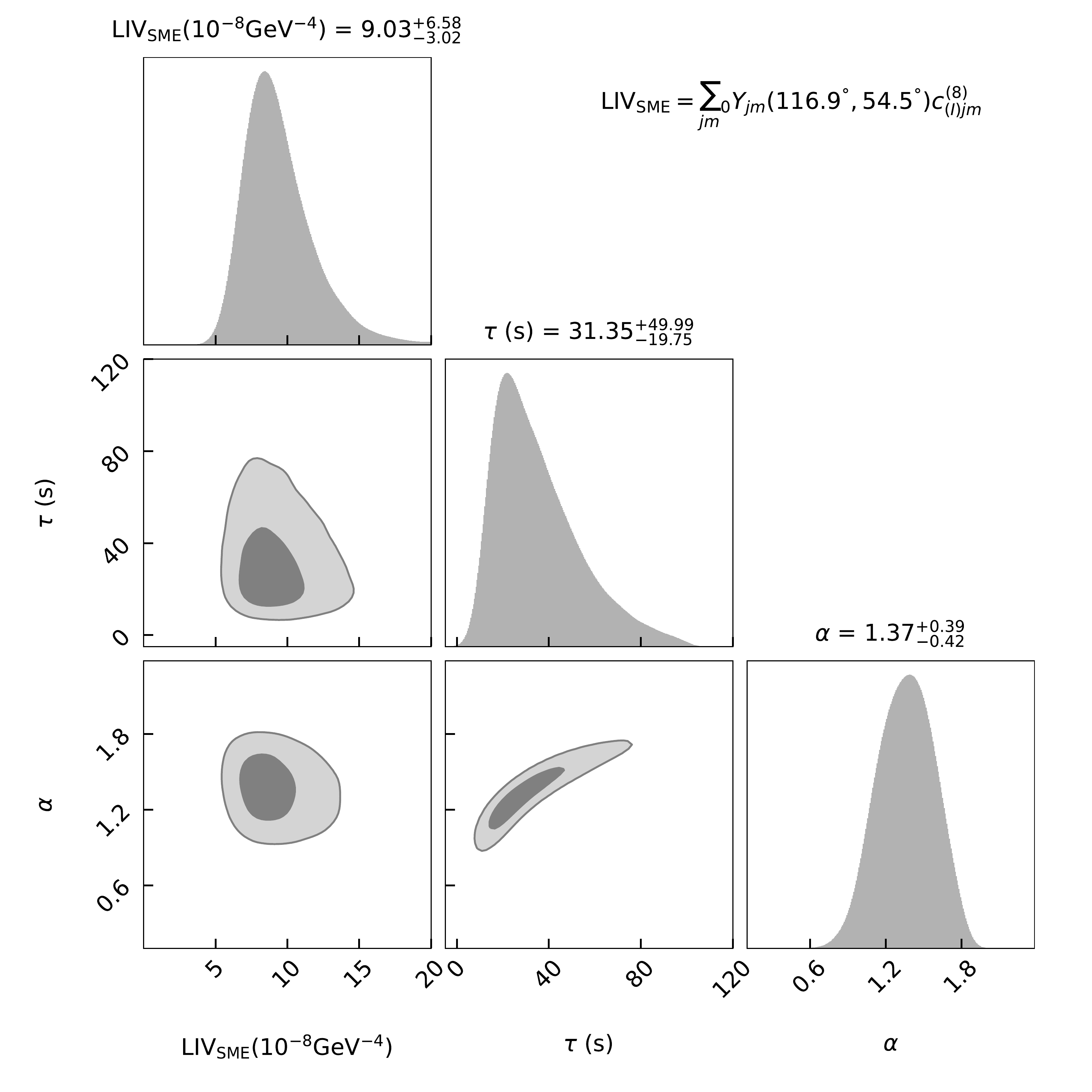}
\caption{Same as Figure~\ref{MyFigC}, but for vacuum dispersion with the standard model extension. Parameter constraints shown on the top and bottom panels correspond to the cases of the mass dimension $d=6$ and $8$, respectively.
}\label{MyFigD}
\end{figure}

With the LIV-induced time delay for standard model extension, the free parameters to be constrained are ${\bm \theta}=\{\sum_{jm} {}_{0}Y_{jm}(\theta, \phi) c_{(I)jm}^{(d)}, \tau, \alpha\}$. Here, we consider the limits of spherical coefficients for LIV effects with taking mass dimension $d=6,8$ in turn. We show the dynamic nested sampling outcomes in Figure~\ref{MyFigD} and in Table~\ref{MyTabB} as well.
We also estimate the constraints on isotropic Lorentz violation for different combination coefficients $c_{(I)00}^{(d)}$ at $2\sigma$ confidence levels.

\subsection{The Statistical Significance of spectral-lag transition}\label{Sec:}
Here, we perform a Bayesian model selection analysis to quantify the statistical significance of the transition from positive lags to negative lags that we found in GRB~190114C. This method quantifies the comparison of two hypothetical models by using the ratio of Bayesian \emph{evidence} (${\mathcal Z}$), the so-called Bayesian odds ratio (or Bayes factor; \citealt{Trotta2008ConPh,Trotta2017arXiv170101467T}). If the \emph{prior} probabilities of the comparing models are equal (this is the case we take into account in the Bayesian analysis), the Bayes factor is given by
\begin{equation}
{\mathcal B}_{\rm 21}\equiv \frac{{\mathcal Z}_2}{{\mathcal Z}_1}=\frac{\int{p({\bm D}_{\rm \tau}|M_2({\bm \theta}_2)})p(M_2({\bm \theta}_2))d {\bm \theta}_2}{\int{p({\bm D}_{\rm \tau}|M_1 ({\bm \theta}_1) })p(M_1 ({\bm \theta}_1) )d {\bm \theta}_1},
\end{equation}
where $M_1$ and $M_2$ correspond to the models without (the null hypothesis) and with the time lags of LIV effects, ${\bm \theta}_1=\{\tau, \alpha\}$, and ${\bm \theta}_2={\bm \theta}$. An empirical qualitative criterion for evaluating the strength of evidence for a preferred model with the Bayes factor is based on the `Jeffreys scale' \citep{Trotta2017arXiv170101467T, Ivezi2019sdmm.book}.
With the effective estimate of \emph{evidence} from the dynamic nested sampling procedure in  Section~\ref{Sec:constraints}, we obtain $\ln{\mathcal B}_{\rm 21,1}=175.08$ and $\ln{\mathcal B}_{\rm 21,2}=174.52$ for the linear and quadratic LIV cases, respectively.
The constraints on SME coefficients yield $\ln{\mathcal B}_{\rm 21}=177.43$ and 173.15 for Lorentz violation with $d=6$ and 8, respectively. 
The limits on the parameters $\tau$ and $\alpha$ in $M_1$ are also shown in Table~\ref{MyTabB}, and $\tau$ has a very large uncertainty. Thus we do not plot this result here. The estimated values of \emph{evidences} indicate that the odds ratios are much greater than 150:1, directing to decisive evidence for the spectral-lag transition of GRB~190114C.


\section{Conclusion and Discussions}\label{Sec:Sec4}
By analyzing the arrival time delays between broad light curves in different energy bands observed in GRB~190114C, we find a well-observed transition from positive lags to negative lags at around 0.7~MeV, analogous to that from GRB~160625B.
This spectral lag behavior allows us to perform a robust test for the LIV effects' possible signatures by directly fitting the detailed spectral lag measurements observed in this astrophysical source. Using the dynamic nested sampling method based on the Bayes' Theorem, the outcomes of model fitting give a suitable formulation for the intrinsic energy-dependent time lag and lower limits on the quantum-gravity energy scales, i.e., $E_{\rm QG,1}\ge2.23\times 10^{14}$~GeV and $E_{\rm QG,2}\ge 0.87\times 10^6$~GeV in linear and quadratic ($n=2$) LIV cases at 2$\sigma$ confidence levels, respectively. Considering a negative spectral lag due to LIV in the SME framework, we obtain $\sum_{jm} {}_{0}Y_{jm}(\theta, \phi) c_{(I)jm}^{(d)}\le 6.77\times10^{-13}$~GeV$^{-2}$ and $1.56\times10^{-7}$~GeV$^{-4}$ (2$\sigma)$ for LIV effects with mass dimensions $d=6$ and 8, respectively.
The Bayes factors obtained in both the LIV models with Taylor expansion and SME and the null hypothesis of no LIV-induced time delay indicate decisive evidence for the existence of spectral-lag transition. 

Because the negative lags exist in the softer energy bands and the lower redshift, the spectral lags of GRB~190114C are not sensitive enough to the LIV constraints compared to GRB~160625B ($z=1.41$). The limits of $E_{\rm QG}$ we obtained from GRB~190114C are below that from GRB~160625B with about one order of magnitude in linear and quadratic LIV models.
The lower limit in the linear LIV model is four orders of magnitude below the Planck scale, one order of magnitude below obtained by \cite{Ellis2006APh} who considered the intrinsic time lag as an unknown constant. The result for the quadratic LIV case is five orders of magnitude below the limit of multi-TeV gamma-ray flare from Mrk 501.
Our constraints on the SME coefficients are somewhat weaker than existing bounds,
but they can be viewed as comparatively robust and have the promise to complement existing LIV constraints.

Despite the current relatively lower constraints, GRBs are advantageous for their detectability at high redshifts with long-lasting VHE emissions and short spectral lags compared to such competitive strong gamma-ray emitters like AGNs.
We are desperate for such kinds of GRB sources emerging at high redshifts with higher energy emissions and higher temporal resolutions for more stringent LIV constraints.
To date, the sub-TeV and even TeV photons have been detected in GRB~180720B ($z=0.653$; \citealt{Abdalla2019Natur.575..464A}) and GRB~190114C \citep{MAGIC2019Natur}, such detections are expected to become routine (see discussions by \citealt{Zhang2019Natur.575..448Z}). The performance of observatories for the TeV astronomy, e.g., the future Cherenkov Telescope Array (CTA, $\sim$30~GeV -- 100~TeV) and the Large High Altitude Air Shower Observatory (LHAASO, $\sim$100~GeV -- 1~PeV) in China, could be rewarding for conducting the LIV constraints from VHE GRBs in the next decades. If only the VHE GRB events are detected, the spectral lag transition may be determined. Our method can release the systematic effects from the assumptions of intrinsic spectral and temporal distributions of VHE emissions beyond the detector's sensitivity and the extragalactic background light model, which severely bias the constraints by searching for an energy dependence in the arrival time of VHE photons \citep{Acciari2020PhRvL.125b1301A}.
On the other hand, the physical origin of the intrinsic time lag is still an issue as the emission mechanism of GRBs remains largely unknown.
\cite{Du2019ApJ} found that the spectral lag is strongly related to the spectral evolution. This gives a possibility to investigate the physical origin of intrinsic time lag with the detailed temporal and spectral features of any GRB, whilst exploring the emission mechanism of the prompt gamma-rays.


\acknowledgments
We are grateful to Da-Bin Lin for his useful discussions, and to Jing Li and Yun Wang for their kind helps.
We acknowledge the use of the public data from the Fermi data archives.
This work is partially supported by the National Natural Science Foundation of China
(grant Nos. 11673068, 11725314, 11533003, U1831122, and 11633001), the Youth Innovation Promotion
Association (2017366), the Key Research Program of Frontier Sciences (grant Nos. QYZDB-SSW-SYS005
and ZDBS-LY-7014), and the Strategic Priority Research Program ``Multi-waveband gravitational wave universe''
(grant No. XDB23000000) of Chinese Academy of Sciences. BBZ acknowledges the supported by the Fundamental Research Funds for the Central Universities (14380035). This work is supported by National Key Research and Development Programs of China (2018YFA0404204), the National Natural Science Foundation of China (Grant Nos. 11833003), and the Program for Innovative Talents, Entrepreneur in Jiangsu.


\clearpage

\clearpage
\begin{table}
\centering \caption{The observed time lags between the lowest energy band ($10-15$~keV)
and any other high energy bands for light curves in intervals $-1$--14~s of GRB~190114C.}\label{MyTabA}
\begin{tabular}{cc}
\hline
\hline
 Energy & lag   \\
 (keV) & (s)  \\
\hline
15-25 & $0.227\pm0.0381$\\
25-35 & $0.542\pm0.071$\\
35-50 & $0.853\pm0.095$\\
50-75 & $0.987\pm 0.142$\\
75-100&$1.002\pm0.127$\\
100-125&$1.016\pm0.122$\\
125-150&$1.024\pm0.113$\\
150-175&$0.999\pm0.101$\\
175-200&$0.968\pm0.086$\\
200-250&$0.981\pm0.081$\\
250-300&$1.023\pm0.082$\\
300-350&$1.023\pm0.081$\\
350-500&$0.943\pm0.081$\\
500-650&$0.938	\pm0.092$\\
650-800&$0.801	\pm0.143$\\
800-1200&$0.634\pm0.069$\\
1200-1500&$0.328\pm0.073$\\
1500-1700&$0.032\pm0.111$\\
1700-5000&$-0.315\pm0.070$\\
\hline
\end{tabular}
\label{MyTabA}
\end{table}

\begin{table}
\begin{center}
\small\addtolength{\tabcolsep}{-3pt}
\centering \caption{The constraint results at 2$\sigma$ confidence levels, the reduced chi-squared, and the Bayes factor for different hypothetical models.}
\begin{tabular}{llcc}
\hline\hline
 Model (Parameters) &  2$\sigma$ bounds & $\chi^2/{\rm dof}$ & Bayes factor \\
\hline
No LIV ($\tau, \alpha$) & $353.42^{+357.66}_{-263.87}$~GeV, $2.46^{+0.28}_{-0.55}$~s & 384.63/17 & Null hypothesis \\
Linear LIV ($\log_{10}E_{\rm QG,1}, \tau, \alpha$) & $14.49^{+0.12}_{-0.13}$~GeV, $10.61^{+22.70}_{-5.52}$~s, $0.84^{+0.50}_{-0.41}$ & 8.39/16 & 1.08$\times10^{76}$ \\
Quadratic LIV ($\log_{10}E_{\rm QG,2}, \tau, \alpha$) & $6.00^{+0.06}_{-0.06}$~GeV,  $22.16^{+45.43}_{-13.77}$~s, $1.21^{+0.46}_{-0.42}$ & 7.36/16 & 6.21$\times10^{75}$ \\
{SME$^{d=6}$ ($\sum_{jm}{_0}Y_{jm}(116.9^{\circ},54.5^{\circ})c^{(6)}_{(I)jm}, \tau, \alpha$)} & $5.05^{+1.72}_{-1.25}\times 10^{-13}$~GeV$^{-2}$, $21.26^{+49.00}_{-13.07}$~s, $1.19^{+0.49}_{-0.41}$ & 5.55/16 & 1.14$\times10^{77}$  \\
{SME$^{d=6}(c^{(6)}_{(I)00})$} & $c^{(6)}_{(I)00}\le 2.40\times 10^{-12}$~GeV$^{-2}$  &-- & --\\
{SME$^{d=8}$ ($\sum_{jm}{_0}Y_{jm}(116.9^{\circ},54.5^{\circ})c^{(8)}_{(I)jm}, \tau, \alpha$)} & $9.03^{+6.58}_{-3.02}\times 10^{-8}$~GeV$^{-4}$, $31.35^{+49.99}_{-19.75}$~s, $1.37^{+0.39}_{-0.42}$ & 10.13/16 & 1.58$\times10^{75}$ \\
{SME$^{d=8}(c^{(8)}_{(I)00})$} & $c^{(8)}_{(I)00}\le 5.53\times 10^{-7}$~GeV$^{-4}$  &-- & --\\
\hline
\end{tabular}
\label{MyTabB}
\end{center}
\end{table}

\clearpage
\bibliography{}

\begin{thebibliography}{}
\expandafter\ifx\csname natexlab\endcsname\relax\def\natexlab#1{#1}\fi
\bibitem[Abdalla et al.(2019)]{Abdalla2019Natur.575..464A} Abdalla, H., Adam, R., Aharonian, F., et al.\ 2019, \nat, 575, 464
\bibitem[Abdalla et al.(2019)]{Abdalla2019ApJ} Abdalla, H., Aharonian, F., Ait Benkhali, F., et al.\ 2019, \apj, 870, 93
\bibitem[Abdo et al.(2009a)]{Abdo2009Sci} Abdo, A.~A., Ackermann, M., Arimoto, M., et al.\ 2009a, Science, 323, 1688
\bibitem[Abdo et al.(2009b)]{Abdo2009Natur} Abdo, A.~A., Ackermann, M., Ajello, M., et al.\ 2009b, \nat, 462, 331
\bibitem[Acciari et al.(2020)]{Acciari2020PhRvL.125b1301A} Acciari, V.~A., Ansoldi, S., Antonelli, L.~A., et al.\ 2020, \prl, 125, 021301
\bibitem[Amelino-Camelia et al.(1997)]{Amelino-Camelia1997IJMPA} Amelino-Camelia, G., Ellis, J., Mavromatos, N.~E., et al.\ 1997, International Journal of Modern Physics A, 12, 607
\bibitem[Amelino-Camelia et al.(1998)]{Amelino-Camelia1998Natur} Amelino-Camelia, G., Ellis, J., Mavromatos, N.~E., et al.\ 1998, \nat, 393, 763
\bibitem[Amelino-Camelia(2013)]{Amelino-Camelia2013LRR} Amelino-Camelia, G.\ 2013, Living Reviews in Relativity, 16, 5
\bibitem[Band(1997)]{Band97ApJ} Band, D.~L.\ 1997, \apj, 486, 928
\bibitem[Boggs et al.(2004)]{Boggs2004ApJ} Boggs, S.~E., Wunderer, C.~B., Hurley, K., et al.\ 2004, \apjl, 611, L77
\bibitem[Chang et al.(2012)]{Chang2012APh} Chang, Z., Jiang, Y., \& Lin, H.-N.\ 2012, Astroparticle Physics, 36, 47
\bibitem[Chen et al.(2005)]{Chen05ApJ} Chen, L., Lou, Y.-Q., Wu, M., et al.\ 2005, \apj, 619, 983
\bibitem[Cheng et al.(1995)]{Cheng95AAP} Cheng, L.~X., Ma, Y.~Q., Cheng, K.~S.,
Lu, T., \&amp; Zhou, Y.~Y.\ 1995, \aap, 300, 746
\bibitem[Colladay \& Kosteleck{\'y}(1997)]{1997PhRvD..55.6760C} Colladay, D. \& Kosteleck{\'y}, V.~A.\ 1997, \prd, 55, 6760
\bibitem[Colladay \& Kosteleck{\'y}(1998)]{1998PhRvD..58k6002C} Colladay, D. \& Kosteleck{\'y}, V.~A.\ 1998, \prd, 58, 116002
\bibitem[Du et al.(2019)]{Du2019ApJ} Du, S.-S., Lin, D.-B., Lu, R.-J., et al.\ 2019, \apj, 882, 115
\bibitem[Ellis et al.(2003)]{Ellis2003A&A} Ellis, J., Mavromatos, N.~E., Nanopoulos, D.~V., et al.\ 2003, \aap, 402, 409
\bibitem[Ellis et al.(2006)]{Ellis2006APh} Ellis, J., Mavromatos, N.~E., Nanopoulos, D.~V., et al.\ 2006, Astroparticle Physics, 25, 402
\bibitem[Ellis \& Mavromatos(2013)]{Ellis2013APh} Ellis, J., \& Mavromatos, N.~E.\ 2013, Astroparticle Physics, 43, 50
\bibitem[Ellis et al.(2019)]{Ellis2019PhRvD} Ellis, J., Konoplich, R., Mavromatos, N.~E., et al.\ 2019, \prd, 99, 083009
\bibitem[Gao et al.(2015)]{Gao2015ApJ} Gao, H., Wu, X.-F., \& M{\'e}sz{\'a}ros, P.\ 2015, \apj, 810, 121
\bibitem[Goldberg et al.(1967)]{Goldberg1967JMP.....8.2155G} Goldberg, J.~N., Macfarlane, A.~J., Newman, E.~T., et al.\ 1967, Journal of Mathematical Physics, 8, 2155
\bibitem[Gropp et al.(2019)]{Gropp2019GCN.23688....1G} Gropp, J.~D., Kennea, J.~A., Klingler, N.~J., et al.\ 2019, GRB Coordinates Network 23688, 1
\bibitem[Hamburg et al.(2019)]{Hamburg2019GCN.23707....1H} Hamburg, R., Veres, P., Meegan, C., et al.\ 2019, GRB Coordinates Network 23707, 1
\bibitem[Ioka \& Nakamura(2001)]{Ioka01ApJL} Ioka, K., \& Nakamura, T.\ 2001, \apjl, 554, L163
\bibitem[Ivezi{\'c} et al.(2019)]{Ivezi2019sdmm.book} Ivezi{\'c}, {\v{Z}}., Connelly, A.~J., Vanderplas, J.~T., et al.\ 2019, Statistics, Data Mining, and Machine Learning in Astronomy, by {\v{Z}}. Ivezi{\'c} et al. ISBN: 9780691198309. Princeton University Press, 2019
\bibitem[Kann et al.(2019)]{Kann2019GCN.23710....1K} Kann, D.~A., Thoene, C.~C., Selsing, J., et al.\ 2019, GRB Coordinates Network 23710, 1
\bibitem[Kocevski et al.(2020)]{Kocevski2020AAS} Kocevski, D., Fermi Gamma-ray Burst Monitor Team, \& Fermi Large Area Telescope Collaboration\ 2020, American Astronomical Society Meeting Abstracts, 434.02
\bibitem[Kosteleck{\'y}(2004)]{2004PhRvD..69j5009K} Kosteleck{\'y}, V.~A.\ 2004, \prd, 69, 105009
\bibitem[Kosteleck{\'y} \& Potting(1995)]{Kosteleck1995PhRvD..51.3923K} Kosteleck{\'y}, V.~A. \& Potting, R.\ 1995, \prd, 51, 3923
\bibitem[Kosteleck{\'y} \& Mewes(2002)]{Kosteleck2002PhRvD..66e6005K} Kosteleck{\'y}, V.~A. \& Mewes, M.\ 2002, \prd, 66, 056005
\bibitem[Kosteleck{\'y} \& Mewes(2008)]{Kosteleck2008ApJ...689L...1K} Kosteleck{\'y}, V.~A. \& Mewes, M.\ 2008, \apjl, 689, L1
\bibitem[Kosteleck{\'y} \& Mewes(2009)]{Kosteleck2009PhRvD..80a5020K} Kosteleck{\'y}, V.~A. \& Mewes, M.\ 2009, \prd, 80, 015020
\bibitem[Krimm et al.(2019)]{Krimm2019GCN.23724....1K} Krimm, H.~A., Barthelmy, S.~D., Cummings, J.~R., et al.\ 2019, GRB Coordinates Network 23724, 1
\bibitem[Lu et al.(2006)]{Lu06MNRAS} Lu, R.-J., Qin, Y.-P., Zhang, Z.-B., \& Yi, T.-F.\ 2006, \mnras, 367, 275
\bibitem[Lu et al.(2018)]{Lu18ApJ} Lu, R.-J., Liang, Y.-F., Lin, D.-B., et al.\ 2018, \apj, 865, 153
\bibitem[MAGIC Collaboration et al.(2019)]{MAGIC2019Natur} MAGIC Collaboration, Acciari, V.~A., Ansoldi, S., et al.\ 2019, \nat, 575, 455
\bibitem[Mattingly(2005)]{Mattingly2005LRR} Mattingly, D.\ 2005, Living Reviews in Relativity, 8, 5
\bibitem[Meegan et al.(2009)]{Meegan2009ApJ} Meegan, C., Lichti, G., Bhat, P.~N., et al.\ 2009, \apj, 702, 791
\bibitem[Nemiroff et al.(2012)]{Nemiroff2012PhRvL} Nemiroff, R.~J., Connolly, R., Holmes, J., et al.\ 2012, \prl, 108, 231103
\bibitem[Newman \& Penrose(1966)]{Newman1966JMP.....7..863N} Newman, E.~T. \& Penrose, R.\ 1966, Journal of Mathematical Physics, 7, 863
\bibitem[Norris et al.(1986)]{Norris86ApJ} Norris, J.~P., Share, G.~H., Messina,
D.~C., et al.\ 1986, \apj, 301, 213
\bibitem[Norris et al.(2000)]{Norris00ApJ} Norris, J.~P., Marani, G.~F., \& Bonnell, J.~T.\ 2000, \apj, 534, 248
\bibitem[Pan et al.(2020)]{Pan2020ApJ} Pan, Y., Qi, J., Cao, S., et al.\ 2020, \apj, 890, 169
\bibitem[Peng et al.(2007)]{Peng07CJAA} Peng, Z.-Y., Lu, R.-J., Qin, Y.-P., \& Zhang, B.-B.\ 2007, \cjaa, 7, 428
\bibitem[Planck Collaboration et al.(2018)]{Planck2018arXiv180706209P} Planck Collaboration, Aghanim, N., Akrami, Y., et al.\ 2018, arXiv e-prints, arXiv:1807.06209
\bibitem[Schaefer(1999)]{Schaefer1999PhRvL} Schaefer, B.~E.\ 1999, \prl, 82, 4964
\bibitem[Selsing et al.(2019)]{Selsing2019GCN.23695....1S} Selsing, J., Fynbo, J.~P.~U., Heintz, K.~E., et al.\ 2019, GRB Coordinates Network 23695, 1
\bibitem[Shao et al.(2017)]{Shao17ApJ} Shao, L., Zhang, B.-B., Wang, F.-R., et al.\ 2017, \apj, 844, 126
\bibitem[Shen et al.(2005)]{Shen05MNRAS} Shen, R.-F., Song, L.-M., \& Li, Z.\ 2005, \mnras, 362, 59
\bibitem[Shenoy et al.(2013)]{Shenoy13ApJ} Shenoy, A., Sonbas, E., Dermer, C., et al.\ 2013, \apj, 778, 3
\bibitem[Speagle(2020)]{Speagle2020MNRAS.493.3132S} Speagle, J.~S.\ 2020, \mnras, 493, 3132
\bibitem[Trotta(2008)]{Trotta2008ConPh} Trotta, R.\ 2008, Contemporary Physics, 49, 71
\bibitem[Trotta(2017)]{Trotta2017arXiv170101467T} Trotta, R.\ 2017, arXiv:1701.01467
\bibitem[Uhm \& Zhang(2016)]{Uhm&Zhang16} Uhm, Z.~L., \& Zhang, B.\ 2016, \apj, 825, 97
\bibitem[Uhm et al.(2018)]{Uhm2018ApJ} Uhm, Z.~L., Zhang, B., \& Racusin, J.\ 2018, \apj, 869, 100
\bibitem[Ukwatta et al.(2010)]{Ukwatta2010ApJ} Ukwatta, T.~N., Stamatikos, M., Dhuga, K.~S., et al.\ 2010, \apj, 711, 1073
\bibitem[Vasileiou et al.(2013)]{Vasileiou2013PhRvD} Vasileiou, V., Jacholkowska, A., Piron, F., et al.\ 2013, \prd, 87, 122001
\bibitem[Wei et al.(2015)]{Wei2015PhRvL} Wei, J.-J., Gao, H., Wu, X.-F., et al.\ 2015, \prl, 115, 261101
\bibitem[Wei et al.(2017a)]{Wei2017ApJL} Wei, J.-J., Zhang, B.-B., Shao, L., et al.\ 2017, \apjl, 834, L13
\bibitem[Wei et al.(2017b)]{Wei2017ApJ...842..115W} Wei, J.-J., Wu, X.-F., Zhang, B.-B., et al.\ 2017, \apj, 842, 115
\bibitem[Xiao \& Ma(2009)]{Xiao2009PhRvD} Xiao, Z., \& Ma, B.-Q.\ 2009, \prd, 80, 116005
\bibitem[Yi et al.(2006)]{Yi06MNRAS} Yi, T., Liang, E., Qin, Y., \& Lu, R.\ 2006, \mnras, 367, 1751
\bibitem[Zhang et al.(2009)]{Zhang09ApJ} Zhang, B., Zhang, B.-B., Virgili, F.~J., et al.\ 2009, \apj, 703, 1696
\bibitem[Zhang(2019)]{Zhang2019Natur.575..448Z} Zhang, B.\ 2019, \nat, 575, 448
\bibitem[Zhang et al.(2012)]{ZhangBB2012ApJ} Zhang, B.-B., Burrows, D.~N., Zhang, B., et al.\ 2012, \apj, 748, 132
\bibitem[Zhang \& Ma(2015)]{Zhang2015APh} Zhang, S., \& Ma, B.-Q.\ 2015, Astroparticle Physics, 61, 108

\end{thebibliography}

\end{document}